\documentclass[12pt,a4paper]{article}

\usepackage[utf8]{inputenc}
\usepackage[T1]{fontenc}
\usepackage{graphicx,amsmath,amssymb, color, stmaryrd}
\usepackage{tikz}
\usetikzlibrary{positioning}
\usepackage{tabularx}
\usepackage{multirow}
\usepackage{enumerate}
\usepackage{float}
\usetikzlibrary{arrows}
\usepackage{natbib}
\bibpunct[,]{(}{)}{,}{a}{}{,}
\usepackage[title]{appendix}
\usepackage{mathtools}
\usepackage{amsthm}
\usepackage[margin=1in]{geometry}
\usepackage{authblk}

\usepackage{fancyhdr,ifoddpage}

\newcommand\keywords{%
\noindent\textbf{Keywords:}\ }

\newtheorem{lemma}{Lemma}
\newtheorem{proposition}{Proposition}
\newtheorem*{proposition*}{Proposition}

\theoremstyle{definition}
\newtheorem{example}{Example}
\theoremstyle{remark}
\newtheorem{remark}{Remark}

\newcommand{\E}{\mathrm{E}}

\title{\bf Linguistic Indirectness in Public Cheap-Talk Games}

\author[1]{Liping Tang}
\author[2]{Michiko Ogaku}
\affil[1]{Institute of Logic and Cognition, Department of Philosophy, Sun Yat-sen University, Guangzhou 510275, China}
\affil[2]{Faculty of Economics, Nagasaki University, 4-2-1 Katafuchi, Nagasaki, 850-8506, Japan, ohgaku@nagasaki-u.ac.jp}
\date{}                     
\begin{document}
\bibliographystyle{plainnat}
\maketitle

\begin{abstract}
We study linguistic indirectness when speakers attend to social ties.  
Social ties are modeled by a graph, and conferences are the sets of nodes that hear a message. 
Conference worth is a distance polynomial on the graph; allocations are given by the Myerson value of the conference-restricted worth, 
which yields the bargaining-power components for each participant. Aggregating these components gives 
an effective bias that,  via a Partition--Threshold rule,  pins down the number of equilibrium message partitions in a cheap talk game. 
Results: (i) among trees, stars maximize worth, leading to weakly fewer equilibrium partitions; (ii) on stars, we derive closed-form effective biases, with a witness-hub marginal effect of adding leaves changing sign at $\delta^{\ast}=0.6$; (iii) for two stars joined by one link, two-star (hub{\footnotesize \textrightarrow}hub) vs big-star (hub{\footnotesize \textrightarrow}leaf) precision flips at $8/15$ for the same number of nodes; private leaf--leaf conferences are most informative.  


\end{abstract}

\section{Introduction} 
We study linguistic indirectness when speakers attend to social relationships. Social ties are given by an undirected graph $G=(V,E)$ and shape word choice between interlocutors---\emph{but do not constrain the routing of information}. 
Their choice of words is determined by the partition equilibria in \cite{Crawford1982}  cheap-talk games. 
In the classic Crawford--Sobel model, there is an exogenously given conflict between interlocutors, which is called a bias $b>0$; 
here, we endogenise it by constructing a TU characteristic-function (point) game $(V,r^{v^G}_{\mathcal{H}})$ from social ties and conference structure $\mathcal{H}$ and apply the Myerson value. 

A conference is 
any subset $C \subset V$ with $|C| \geq 2$ whose members meet and hear the message \citep{myerson1980conference, van1992allocation}. 
For any $C \subset V$, the conference worth is the distance polynomial 
\[
v^G(C)=2 \sum_{t \geq 1} U_t(G[C]) \delta^t, \ \delta: \text{link weight}.
\]
We allocate conference worth via the Myerson value applied to the conference-restricted worth,
\[
\mu(V;r^{v^G}_{\mathcal{H}}), \ r^{v^G}_{\mathcal{H}} (C) = \sum_{B \in C / \mathcal{H}} v^G(B).
\]
The resulting cardinal indices serve as bargaining-power components: the bargaining power \emph{of $i$ over $j$} at a conference is
\[
b^j_i := \mu_j(V; r^{v^G}_{\mathcal{H}}) - \mu_j(V \setminus \{i\}; r^{v^G}_{\mathcal{H}|_{V \setminus \{i\}}}).
\]
Aggregating these bargaining-power components yields an effective bias $b_{\mathrm{eff}}$, which, via a Partition--Threshold (PT) rule, pins down the number of equilibrium message partitions. 
The cheap-talk environment is NTU (no side payments), while $(V,r^{v_G}_{\mathcal{H}})$ is a TU characteristic-function game used only as a cardinal \emph{index} of bargaining power; no transfers occur in equilibrium play.

Our contribution is threefold. (i) \emph{Stars maximize} $v^G(\cdot)$ among trees of fixed size; hence stars yield weakly fewer equilibrium partitions than any other tree (PT). (ii) On stars, we obtain \emph{closed-form} Myerson value (hub/leaf/witness) and the implied $b_{\mathrm{eff}}$; in the witness-hub case, the marginal effect of adding leaves changes sign at $\delta^{\ast}=0.6$. (iii) For two stars joined by a single link, we compare protocols: the two-star (hub $\to$ hub) vs the big-star (hub $\to$ leaf) with the same number of nodes, equilibrium message precision flips at $\delta=8/15$; the most informative communication arises when the hubs ask their leaves to speak privately in the smallest conference (the pair only). 

Our evaluation basis---\emph{distance polynomial + Myerson on a conference}---isolates how shifting distance mass from long to short paths raises $b_{\mathrm{eff}}$ and reduces the number of partitions of equilibrium messages. We relate to the graph-/conference-restricted TU game, network formation with pairwise funding \citep{jackson1996strategic, bloch2006definitions}, and the pragmatics of indirectness \citep{pinker2008logic,brown1987politeness,goffman1955face,tang2024linguistic}. 

Section \ref{sec:model} sets up primitives, restricted worth, $b_{\mathrm{eff}}$, and PT.   
Section \ref{sec:results} analyzes single stars. Section \ref{sec:two-stars} studies two-star joins and protocol comparisons. 
Section \ref{sec:conclusion} concludes; proofs are in the appendix. 

\section{Basic model}\label{sec:model} 
\subsection{Primitives and conventions}
Let $G=(V,E)$ be an undirected graph with $|V|=n$ players. 
Edges of the base graph $G$ represent social relationships that players care about. They are not message routing constraints. 
Communication is governed by a conference structure $\mathcal{H}$ defined below. 
Let $\mathcal{H} \subset \{H \in 2^{\scriptscriptstyle V} : |H| \geq 2\}$ be a conference structure (hyperedges) and let $(V,\mathcal{H})$ be an (undirected) hypergraph representing the communication possibilities. 
The triple $(V,v,\mathcal{H})$ is referred to as a communication situation, where $v$ is a value function defined on $2^{\scriptscriptstyle V}$. We specify a value function $v^G$ in the next subsection.  
Let $\mathrm{HCS}^{\scriptscriptstyle V}$ be the class of all hypergraph communication situations. 
If $(V,v,\mathcal{H}) \in HCS^{\scriptscriptstyle V}$, then a coalition $S \subset V$ can coordinate communication in the conferences 
in $\mathcal{H}(S):=\{H \in \mathcal{H}  :  H \subset S\}$. 
Given $\mathcal{H}(S)$, players $i$ and $j$ are connected by $\mathcal{H}(S)$ if $i = j$ or there exists some sequence of conferences 
$(S_1, \cdots, S_m)$ such that 
\begin{align*}
&i \in S_1, \ j \in S_m, \ \{S_1,\cdots, S_m\} \subset \mathcal{H}(S), \ \text{and } \\
&S_k \cap S_{k+1} \ne \varnothing \ \text{ for every }k=1,\cdots,m-1.
\end{align*}
Let $S / \mathcal{H}$ denote the partition of $S$ defined by this connectedness relation:
\[
S / \mathcal{H}=\big\{
\{
j  :  \text{$i$ and $j$ are connected by }\mathcal{H}(S) 
\}  :  i \in S
\big\}.
\]
Once a conference $C \in S / \mathcal{H}$ is open, all members of $C$ can potentially communicate, regardless of paths (social relations) in $G$. 
We study indirect expressions in a conference $C$  
on a hypergraph communication situation $(V, v^G, \mathcal{H})$ with 
\begin{align}
\mathcal{H}=\big\{
\{i,j\} : \{i,j\} \in E
\big\}. \label{eq:H-3-node}
\end{align}
Unless stated otherwise, we use \eqref{eq:H-3-node}; in Section \ref{sec:two-stars} we modify $\mathcal{H}$ to encode cross-block communication protocols. 

Given a conference $C$, a sender $S$, receiver $R$, and witness set $W \subset C \setminus \{S,R\}$ play the Crawford--Sobel cheap-talk game: $S$ observes state 
$\theta \in [0,1]$, sends message $m$, $R$ selects action $a$, and costs (payoffs) are quadratic with intrinsic bias $b>0$ between $S$ and $R$. 
\begin{align*}
&\text{Cost of }R: \quad -(a-\theta)^2, \\
&\text{Cost of }S: \quad -(a-(\theta +b))^2.
\end{align*}
In the classic Crawford--Sobel model, $b$ is exogenous; here, our graph (social relationship) restricted game endogenises it: with a single dyadic link, the Myerson value on a conference we mention below assigns $\delta$. Equilibria of our interest are partition equilibria, where the sender's message will be an interval of a partition of $[0,1]$ which contains the realisation of $\theta$; \textbf{fewer partitions mean a more indirect message}. 

Each player also cares about witnesses' opinions, weighted by the Myerson value on the conference $\mu_i(V;r^{v^G}_{\mathcal{H}})$---the Shapley value applied to the components 
that form as players join the conference \citep{myerson1977graphs}. 

\subsection{Distance polynomial worth $v^G$}
Let $G[S]$ be the subgraph induced by $S \subseteq V$. 
For $i,j\in S \subset V$, let $t_{ij}(G[S])$ be the shortest-path distance
in $G[S]$, with $t_{ij}(G[S])=\infty$ if $i$ and $j$ are disconnected, and fix a link/value parameter $\delta\in(0,1)$.

We define a value function $v^G$ as a distance polynomial:
\[
  v^G(S)\;:=\;\sum_{i\in S}\;\sum_{\substack{j\in S \\ j\ne i}} \delta^{\,t_{ij}(G[S])} \quad \forall S \subset V, 
  \quad\text{with }\;\delta^{\infty}:=0.
\]
Group by distance $k\in\{1,2,\dots\}$: 
\[v^G(S)=2\sum_{k\ge 1} U_k(G[S])\;\delta^{k} \quad \forall S \subset V,\]
where $U_k(G[S])$ = \#\{unordered pairs at distance $k$\}.
The distance histogram $\{U_k(G[S])\}$ is evaluated at $\delta$. 

For a star with $k$ leaves:
\[
U_1(S_k)=k,\qquad U_2(S_k)=\binom{k}{2}=\tfrac{k(k-1)}{2}.
\]
$\delta$-weighted distance contributions for $S_k$ are given by: $V_1=2k\delta$, $V_2=k(k-1)\delta^2$; with $\delta \in (0,1)$
\[V_1 \geq V_2 \Longleftrightarrow \delta \leq \frac{2}{k-1} \Longrightarrow \text{ for $\delta$ close to }1, V_1<V_2 \text{ for }k \geq 4.\]

We care about weighted contributions; even when distance-2 pairs are more numerous, the $\delta$-weights can make distance-1 dominate. 

\subsection{Myerson value on graphs and on conferences}
The \emph{Myerson value} of player $i$ in $(G,v^G)$ is defined such that 
\[\mu_{i}(V; v^G)=\sum_{S \subset V \setminus \{i\}} (v^G(S \cup \{i\})-v^G(S)) \left( \frac{s!(n-s-1)!}{n!}\right), \]
where $s=|S|$. 
For $G=S_k$, the Myerson values differ depending on whether the player (node) is the hub or a leaf: 
\[
    \mu_{\text{leaf}}
      = \delta + \frac{2(k-1)}{3}\delta^{2}, 
\quad
    \mu_{\text{hub}}
      = k\delta + \frac{k(k-1)}{3}\delta^{2}.
\]
\begin{remark}
We can interpret that the coefficient of $\delta$ is the distance-1 effect, and the coefficient of $\delta^2$ is the distance-2 effect. 
\end{remark}

One does not need to calculate the heavy sum to obtain the Myerson value $\mu_{i}(G; v^G)$. 
On trees, the Myerson value has a simple path-sharing rule as follows. 
\begin{lemma}[Tree path-sharing rule]\label{lem:tree-path-sharing-rule}
Let $T$ be a tree and fix an ordered pair $(p,q)$ whose unique path $P=\{x_0=p,x_1,\cdots, x_t=q\}$ has length $t$. 
Define the pair-connection unanimity game on $T$ by 
\[
v_{pq}(S)=
\begin{cases}
\delta^{t},& \text{if }P\subseteq S,\\
0,& \text{otherwise.}
\end{cases}
\]
If an allocation rule $\phi$ satisfies Component Efficiency (CE) and Equal Bargaining Power (F) (Myerson's axioms), then 
necessarily,
\begin{align}
\phi_{x}(T,v_{pq})=\begin{cases}
\delta^t/(t+1), & x \in P, \\
0, & x \not\in P. \label{eq:equal-split}
\end{cases}
\end{align}
Conversely, any rule $\phi$ that satisfies \eqref{eq:equal-split} for every pair $p,q$ on $T$ satisfies CE and F. 
\end{lemma}
\begin{remark}
If $G$ is tree, our value function is 
\[
v^G(S) = \sum_p\sum_{q \ne p} v_{pq}(S), \quad \forall S \subset V
\] and the Myerson value of $i$ is given by 
\[
\mu_i(V;v^G) = \sum_{p \ne q} \phi_i(G; v_{pq}).
\]
\end{remark}

A restricted worth (conference) $r^{v^G}:2^{\scriptscriptstyle V} \times 2^{\scriptscriptstyle\mathcal{H}} \to \mathbb{R}$ maps a coalition $C \subset V$ and a set of conferences $\mathcal{A} \subset \mathcal{H}$ to the worth the coalition can achieve when effecting communication in all conferences of $\mathcal{A}(C)$. 
\[
r^{v^G}(C, \mathcal{A}): = \sum_{B \in C / \mathcal{A}} v^G(B) \quad \forall \mathcal{A} \subset \mathcal{H}, \ \forall C \subset V.
\]
The Myerson value on hypergraph communication situations is defined such that 
\[
\mu(V,v^G,\mathcal{H}):= \mu(V;r^{v^G}_{\mathcal{H}}), \ r^{v^G}_{\mathcal{H}}(C):= r^{v^G}(C, \mathcal{H}) \quad \forall C \subset V.
\] 
\begin{example}
(3-node star, $S$--$R$--$W$). Let 
\begin{align*}
\mathcal{H}=\{\{S,R\}, \{R,W\}\}. 
\end{align*}
Then we have  
\[
r^{v^G}_{\mathcal{H}}(\{S,R,W\}) =2 \delta^2 +4 \delta, \
r^{v^G}_{\mathcal{H}}(\{S,R\}) =r^{v^G}_{\mathcal{H}}(\{R,W\})=2 \delta,
\]
and 
\begin{align*}
\mu_{S}(V,v^G, \mathcal{H}) &=\mu_{W}(V,v^G,\mathcal{H})=\delta+ \frac{2}{3}\delta^2, \\
\mu_{R}(V,v^G,\mathcal{H}) &= 2 \delta + \frac{2}{3}\delta^2.
\end{align*}
\end{example}

Adding $\{S,W\}$ to $\mathcal{H}$ does not change any Myerson value because $G[\{S,W\}]$ has no edge.  
\emph{$\mathcal{H}$ governs who can coordinate; worth is always evaluated on induced subgraphs of $G$} (social relation).

\subsection{Bias components and effective bias}
We define the bargaining power of $i$ to $j$ as follows.  
\[
b^j_i := \mu_j(V;r^{v^G}_{\mathcal{H}}) \;-\; \mu_j\!\big(V\setminus\{i\};r^{v^G}_{\mathcal{H}|_{V \setminus \{i\}}}\big), \ 
\mathcal{H}_{\vert X}:=\{H \in \mathcal{H} : H \subset X\}.
\] 
(Here $b_i^j$ measures how much $j$ loses if $i$ is removed; we therefore speak of the bargaining-power
component “from $j$ to $i$,” i.e., $j$’s concern about $i$.) 
These components aggregate into  $b_{\mathrm{eff}}$ via \eqref{eq:eff}.

In our extended cheap-talk game, the costs (payoffs) in the presence of witnesses will be as follows. 
\begin{align*}
&\text{Cost of }R: \quad -(a-\theta)^2, \\
&\text{Cost of }i \in W: \quad -(a-(\theta+b^i_ R))^2,\\
&\text{Cost of }S: \quad -(a-(\theta+b^S_R))^2-\sum_{w \in W}(a-(\theta + b^S_w+ b^w_R))^2. 
\end{align*}
We will derive the Pareto-superior equilibrium of this form of game.

Let $N$ be an integer and a partition of size $N$
\[0 = t_0(N)<t_1(N)<\cdots<t_{N-1}(N)<t_N(N)=1.\]
Consider the strategy of S such that 
\begin{align*}
\sigma(\theta) =\begin{cases}
m_k & \text{ if } \theta \in [t_{k-1}(N),t_k(N)) \\
m_N & \text{ if } \theta \in [t_{N-1}(N),1],
\end{cases}
\end{align*}
with $m_1<\cdots<m_N$. Receiving the message $m_k$, 
the best response of R is $\E[\theta \vert m_k]=(t_{k-1}(N)+t_k(N))/2$. The best response of S with the strategy $\sigma$ 
is shown by Crawford--Sobel. 
Let $\beta(N):=\frac{1}{2N(N+1)}$.         
\begin{proposition*}[Crawford-Sobel]
Crawford-Sobel show that for all $b$ with $\beta(N) \leq b <\beta(N-1)$, the Pareto-superior equilibrium has $N$ partitions.
\end{proposition*}

We show the Partition Threshold Rule in the presence of witnesses. 
\begin{proposition}[Partition--Threshold Rule (PT)]\label{prop:presence-of-witnesses}
In the presence of $|W|$ witnesses, for any effective bias $b_{\mathrm{eff}}$ satisfying \eqref{eq:eff}, 
the Pareto-superior equilibrium has $N$ partitions. 
\begin{align}
\beta(N)\leq  b_{\mathrm{eff}}  < \beta(N-1), \quad b_{\mathrm{eff}} :=\frac{b^S_R+\sum_{w \in W}( b^S_w+b^w_R)}{|W|+1}.  \label{eq:eff}
\end{align}
\end{proposition}
\noindent
The proof is given in the Appendix. 
\vspace{12pt}

\noindent 
Hence, larger $b_{\mathrm{eff}}$ moves the partition count $N$ down by the Crawford--Sobel partition equilibrium thresholds $\beta(N)=\frac{1}{2N(N+1)}$.

\begin{remark}
Equilibria with larger $N$ are more efficient in costs, and the message is more explicit. 
\end{remark}

\begin{example}
(conference of 3-node star, $S-R-W$). With link value $\delta$, $b^S_R=b^W_R=\delta+\frac{2}{3} \delta^2$, $b^S_W=\frac{2}{3}\delta^2$, and we see 
$b_{\mathrm{eff}} = \delta +\delta^2 > \delta$ (two-player conference bias).
 Hence, $N$ weakly falls relative to the benchmark by the $\beta(N)$ thresholds. 
\end{example}

\begin{remark}
The effective bias is a simple distance polynomial via Myerson splits. 
It tells us the number of partitions in the Pareto-superior equilibrium by the Partition--Threshold Rule (PT).
That is, once the effective bias moves, PT tells us the number of partition moves in the opposite direction. 
\end{remark}

\section{Witness effect in stars} \label{sec:results}
Consider $G$ containing a star $S_k$. Consider a communication situation $(V,v^G, \mathcal{H})$ with $\mathcal{H}$ in \eqref{eq:H-3-node}. 

\begin{lemma}[Stars vs. Trees]\label{lem:star-max}
Fix $n \geq 2$ and $\delta \in (0,1)$. Among all trees $T=(V_T,E_T)$ on $n$ nodes, the star $S_{n-1}$ 
 maximize the value $v^G(V_T)=2\sum_{t\geq 1}U_t(T) \delta^{t}$, where $U_t(T)$ counts unordered node pairs at distance $t$. 
\end{lemma}

Consider a 3-node star $S_2$, where $S=$ sender, $R=$ receiver, and $W=$ witness. 
Consider a conference 
$C=\{S,R,W\}$ and communication possibilities $(V, \mathcal{H})$ with \eqref{eq:H-3-node}. The Myerson values on the conference are given by 
\[
\mu_{\mathrm{hub}}(S_2;r^{v^G}_{\mathcal{H}})=2 \delta + \frac{2}{3}\delta^2, \quad \mu_{\mathrm{leaf}}(S_2; r^{v^G}_{\mathcal{H}})=\delta + \frac{2}{3}\delta^2.
\]
It is easy to see that in every permutation of network roles, from \eqref{eq:eff} the effective bias of $S_3$ is written as 
\[
b_{\mathrm{eff}}=\frac{b^S_R+b^W_R+b^S_W}{2} = \delta+\delta^2 >\delta = b^S_R,
\]
and we have the following proposition. 
\begin{proposition}[Single-witness effect]\label{prop:single-witness}
In any three-node star conference, the presence of a witness weakly \emph{reduces}
the number of equilibrium partitions (hence it raises indirectness) for every permutation of
network roles.
\end{proposition}
\noindent
The proof is given in the Appendix. 

However, when $|W| \geq 2$, the effective bias is role-dependent. 
When the \emph{sender} (or \emph{receiver}) occupies the hub, the effective bias in \eqref{eq:eff} for the conference of $S_k$ is given by 
\begin{align}
 b_{\mathrm{eff}}^{\mathrm{(\text{s.hub})}}
  = \delta\;+\;\underbrace{
      \frac{2(k+1)(k-1)}{3k}\,\delta^{2}
    }_{\textstyle f(k,\delta)}, \label{eq:eff_s_hub}
\end{align}
 strictly increasing in $k$  for all $\delta>0$.
Hence $f(k,\delta)>0$
and $\partial f/\partial k>0$.

\begin{proposition}[Many witnesses on a star] \label{prop:SorR-hub-case}
If $S$ or $R$ is the hub of an $n$-node star, the number of equilibrium partitions in its conference is
non-increasing in $n$; marginal change diminishes as $n\!\to\!\infty$.
\end{proposition}
\noindent
The proof is given in the Appendix.

When a \emph{witness} occupies the hub, the effective bias in \eqref{eq:eff} for the conference of $S_k$ is given by 
\[
  b_{\mathrm{eff}}^{\mathrm{(\text{w.hub})}}
  \;=\; 
\dfrac{2\delta+(4k-5)\tfrac23\delta^{2}}{\,k\,}=: g(k,\delta), 
\]
and see a threshold at $\delta=\delta^*=0.6$.

\begin{proposition}[Peripheral conversations] \label{prop:W-hub-case}
If the hub is a witness and $\delta <\delta^{\ast}=0.6$, then the number of equilibrium partitions in its global conference \emph{increases} (speech becomes more direct).
If $\delta \geq \delta^{\ast}$, the number of equilibrium partitions is non-increasing in $n$. 
For $\delta \in (0,1)$, the marginal effect of adding a witness tends to $0$. 
\end{proposition}
\noindent
The proof is given in the Appendix.

\begin{remark}~

Propositions \ref{prop:single-witness} and \ref{prop:SorR-hub-case} 
show similar effects of audience size in \cite{latane1981psychology}, where audience size influences an individual's stage fright behavior, and its marginal effect diminishes. 

Proposition \ref{prop:W-hub-case} suggests a novel hypothesis generated by our model. 
Latan{\'e}'s theory implies that as the audience size increases, so does the social impact on an individual, assuming that audience characteristics (e.g., attention and interest) remain constant. However, one can imagine a scenario where a large, inattentive audience functions as noise, effectively allowing interlocutors to hide in the crowd. Further research is needed to explore this possibility.
\end{remark}

\section{Communication between two stars}\label{sec:two-stars}
Consider $G$ with two stars $S_k$ and $S_\ell$ on disjoint node sets. 
What manner of speech would be used when the two stars interact? 
Let $E_k$ and $E_\ell$ be the sets of edges in $S_k$ and $S_\ell$, respectively. Similarly, let $V_k$ and $V_\ell$ be the sets of nodes in $S_k$ and $S_\ell$, respectively. 
Let $S_k\text{--}S_\ell$ denote the tree obtained by adding an edge $e$ between, for example, the two hubs. 
The potential link $e$ is a cooperative move that turns the network $G$ into $G+e$ (hence the game $v^G$ into $v^{G+e}$). 
During the negotiation phase, we assume that conferences are held on $G+e$.
Consider a conference structure as follows. 
\[
\mathcal{H}=\mathcal{H}_{\mathrm{intra}} \cup \mathcal{H}_{\mathrm{cross}},
\]
where $\mathcal{H}_{\mathrm{intra}}$ is the union of conference structures of each blocks
\begin{align*}
\mathcal{H}_{\mathrm{intra}}&=\big\{
\{i,j\} : \{i,j\} \in E_k \big\} \cup \big\{ \{i,j\} : \{i,j\} \in E_\ell
\big\}, 
\end{align*}
and $\mathcal{H}_{\mathrm{cross}}$ is the set of pairs sharing the potential link $e$. If the link is made between the hubs, 
\begin{align*}
\mathcal{H}_{\mathrm{cross}}&=\big\{
\{h_k,h_\ell\} \big\}.
\end{align*}
Let $G_0=(V_0,E_0)$ be a graph containing only one star, and $\mathcal{H}_0=\{\{i,j\} : \{i,j\} \in E_0\}$.  
First, compare two conferences with the same number of nodes $n=k+\ell +2$: 
\begin{itemize}
\item[(i)] on $S_k\text{--}S_\ell$, the hub of $S_k$ (sender) speaks to the hub of $S_\ell$ (receiver);
\item[(ii)] on $S_{k+\ell+1}$, the hub speaks to one of its leaves. 
\end{itemize}
\begin{proposition}[Two stars vs. one larger star]\label{prop:two-stars}
Let $m=k+\ell+1$ and $\delta\in(0,1)$. Then
\[
b_{\mathrm{eff}}\big(S_m\text{ (hub}\to\text{leaf)}\big)
-
b_{\mathrm{eff}}\big(S_k\text{--}S_{\ell}\text{ (hub}\to\text{hub)}\big)
=\frac{k\ell\,\delta^2}{m}\Big(\frac{4}{3}-\frac{5}{2}\delta\Big).
\]
Hence, by PT, the big star has weakly fewer equilibrium partitions iff $\delta \leq \delta^{\ast}:=\tfrac{8}{15}$, 
the two-star join has weakly fewer equilibrium partitions iff $\delta\geq \delta^{\ast}$. 
\end{proposition}
\noindent
The proof is given in the Appendix. 

The hubs of $S_k$ and $S_{\ell}$ might find that it is impossible to achieve a Pareto-superior to a 
babbling equilibrium. 
Alternatively, they might consider communicating in a smaller conference $\{h_k,h_\ell\}$, where they do not have to pay attention to the reactions of witness nodes. 
Consider a conference $\{h_k,h_\ell\}$. Then, the effective bias would be reduced to 
\[
b^S_R\big(S_k\text{--}S_{\ell}\text{ (hub}\to\text{hub)}\big)
= \delta+\frac{2}{3}(k+\ell)\delta^2  + \frac{k \ell}{2} \delta^3, 
\] and we see the effective bias would be strictly greater than that in the conference $\{h_m,c\}$, $c \in V_m$ in the big star:
\[
b^S_R\big(S_m\text{ (hub}\to\text{leaf)}\big) = \delta+\frac{2}{3}(k+\ell)\delta^2.
\] 
However, communication in the smaller conference $\{h_k,h_\ell\}$ is much more explicit.
Furthermore, the hubs might consider asking two leaves to communicate to mediate the formation of the hub--hub link. The effective bias in the conference $\{h_k,a\}$, $a \in V_k$, would be 
\[
b^S_R \big(S_k\text{--}S_{\ell}\text{ (hub}\to\text{leaf)}\big)= \delta + \frac{2k}{3} \delta^2 + \frac{\ell}{2}\delta^3,
\]
and the effective bias in the conference $\{a,b\}$, $b \in V_\ell$ would be 
\[
b^S_R \big(S_k\text{--}S_{\ell}\text{ (leaf}\to\text{leaf)}\big)=\frac{1}{2}\delta^3.
\]
The effective bias is smaller whenever $\delta>0$ (since the added terms are positive). So communication will be more explicit. 

Lastly, consider the case where two stars $S_k$ and $S_\ell$ form a leaf--leaf link, where 
\[
\mathcal{H}_{\mathrm{cross}}=\{\{a,b\}\}, \ a \in V_j, b \in V_\ell.
\]
Note that the hub--hub link would provide a higher value than a leaf--leaf link.  
\begin{lemma}\label{lem:hub-hub-vs-leaf-leaf}
Let $k, \ell \geq 1$ and $\delta \in (0,1)$.  
\[
v^{G+e}\big(S_k\text{--}S_{\ell}\text{ (hub}\to\text{hub)}\big)>v^{G+e}\big(S_k\text{--}S_{\ell}\text{ (leaf}\to\text{leaf)}\big).
\]
\end{lemma}
\noindent
The proof is given in the Appendix. 

However, as we see in the following proposition, in most cases, more informative communication is expected in a leaf--leaf link than in a hub--hub link by PT. 

\begin{proposition}\label{prop:hub-hub-vs-leaf-leaf}
Let $k, \ell \geq 1$ and $\delta \in (0,1)$. Define the difference between effective biases in the conference $C_1=V_k \cup V_\ell$ such that 
\[
\Delta(\delta):=b_{\mathrm{eff}}\big(S_k\text{--}S_{\ell}\text{ (hub}\to\text{hub)}\big)
-
b_{\mathrm{eff}}\big(S_k\text{--}S_{\ell}\text{ (leaf}\to\text{leaf)}\big).
\]
Then $\Delta(\delta) \geq 0$ for all $\delta \in (0,1)$ (with equality only at $(k,\ell) =(1,1)$), except when 
$(k,l)=(2,2)$, where $\Delta(\delta)$ changes sign at $\delta_c \doteqdot 0.9949$ and becomes negative for $\delta>\delta_c$.
\end{proposition}
\noindent
The proof is given in the Appendix. 
\begin{remark}
At $\delta=1$ every path of length $t$ contributes $1/(t+1)$. 
In the case $(k,\ell)=(2,2)$, the leaf--leaf link forms a line, while the hub--hub link forms an edge-hubbed star.
The effective bias of the public conversation on the line could be larger at $\delta$ close to one, because 
the two nodes closest to the endpoints of the line have larger bargaining power against the interlocutors. They are the 
ex-hubs of the stars. 
\begin{figure}[H]
\begin{center}\hspace{-24pt}
\includegraphics[width=7cm]{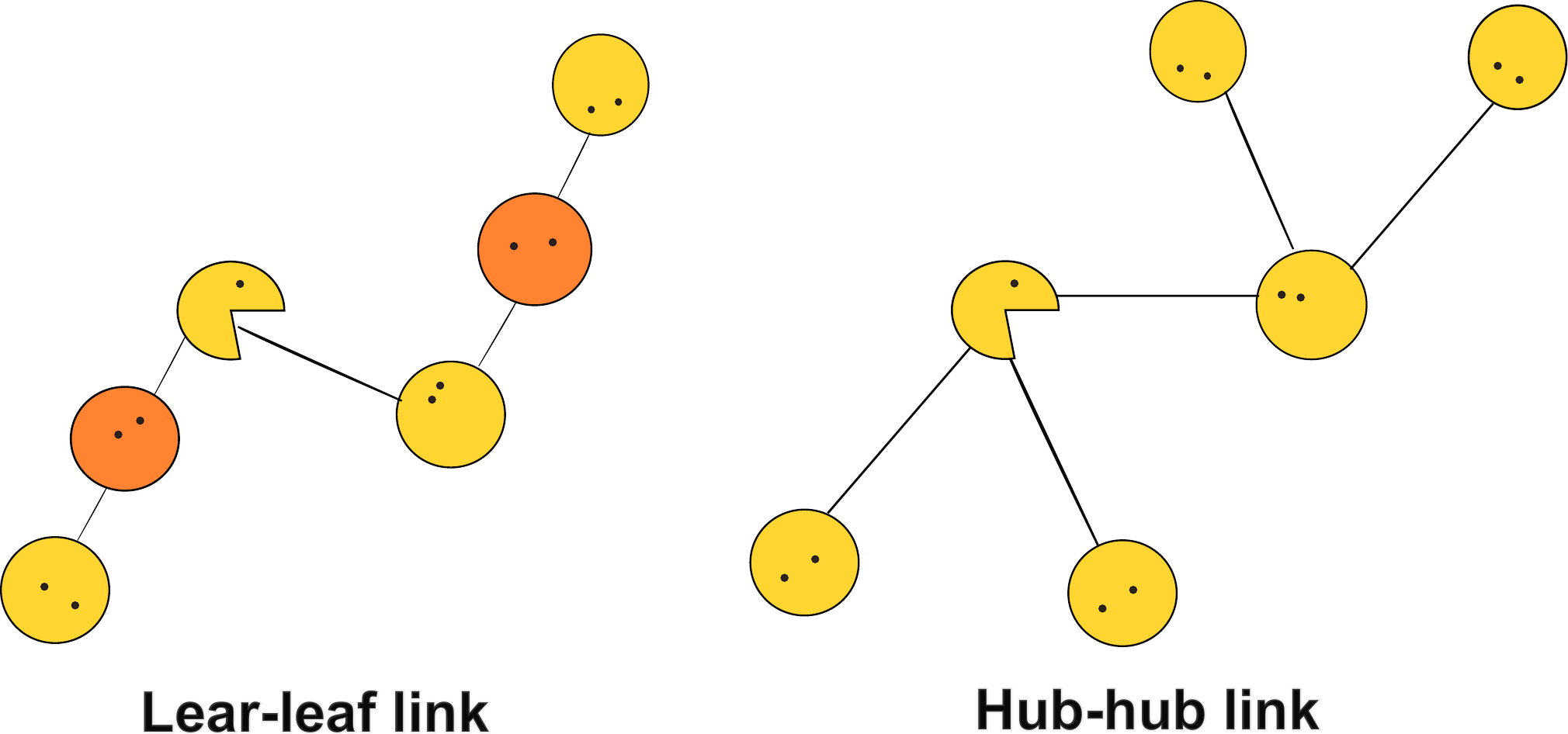} 
\label{fig:star-networks}
\end{center}
\end{figure}
Let $w_{h_k}$ be the ex-hub of $S_k$ in the leaf--leaf link coalition and $w_a$ be an ex-leaf of $S_k$ in the hub--hub link coalition. Then, we have
\[
b^{w_{h_k}}_R =\frac{2}{3}\delta^2 + \delta^3 + \frac{4}{5}\delta^4 + \frac{1}{3}\delta^5 > b^{w_a}_R=\frac{2}{3}\delta^2+\delta^3, 
\] and 
\[
b^S_{w_{h_k}} = \delta + \frac{4}{3} \delta^2 + \delta^3 + \frac{4}{5}\delta^4 + \frac{1}{3}\delta^5 > b^S_{w_a}=\delta+\frac{4}{3}\delta^2+\delta^3.
\]
\end{remark}

\section{Conclusion}\label{sec:conclusion}
The paper explains the use of linguistic indirectness only with graphs and conferences. 
Graphs represent social relationships, and the choice of words is determined using partition equilibria in a cheap talk game. 
The determinant of word choices is the bargaining power of conference participants, which is measured using a distance polynomial and split via the Myerson value. The results are combinatorial. The use of linguistic indirectness is a consequence of distance counting and conference structure.

\bibliography{jecon_ref_lm}

\begin{appendices}
\counterwithin{theorem}{section}
\counterwithin{lemma}{section}
\counterwithin{proposition}{section}
\numberwithin{equation}{section}

\section{Proofs}
\subsection*{Proof of Lemma \ref{lem:tree-path-sharing-rule}}
\begin{proof}
Consider the graph-restricted game on $T$ induced by $v_{pq}$. 
By construction of $v_{pq}$, $S \subset T$ yields value $\delta^t$ iff it contains the entire path $P$; thus, the restricted characteristic function is the unanimity game on the player set $P$.  

\vspace{12pt}

\noindent
(i) Players off $P$ are null. 
By CE and F,  
any player $x \not\in P$ never affects the allocation, so $\phi_x=0$.

\vspace{12pt}

\noindent
(ii) Equal split on $P$. The $t+1$ players are symmetric in the restricted unanimity game.  
By F they must receive equal pay, and by CE their total equals $\delta^t$. 
It follows that the only feasible allocation is the equal split $\delta^t/(t+1)$ on $P$. 

\vspace{12pt}

\noindent
(iii) Sufficiency. Conversely, assume that $\phi$ satisfies \eqref{eq:equal-split} for every $(p,q)$. 
Then, on each connected component, the payoffs sum to the component's worth (CE) and symmetric 
players obtain equal pay (F).
\end{proof}

\subsection*{Proof of Proposition \ref{prop:presence-of-witnesses}}
\begin{proof}
Consider the conditions for a partition equilibrium of size $N$. Let $t(N)=(t_0(N),\dots,t_N(N))$ denote a partition of $[0,1]$ such that $0=t_0(N)<t_1(N)<\cdots<t_N(N)=1$. We write the strategy of S such that for $k=0,\dots,N-1$ 
\[ \mu(\theta, N) = m_k  \quad \text{if }\theta \in (t_{k}(N),t_{k+1}(N)], \]
where $m_0<\cdots<m_{N-1}$. Then the best response of R to each $m_k, \ k=0,\dots,N-1$ is given by
\[a_k=\frac{t_{k-1}(N)+t_{k}(N)}{2}.\]
Given this strategy of R, if $\mu$ is the best response for S, then 
\begin{align}
u(a_k,t_k(N))=u(a_{k+1},t_{k}(N)) \label{eq:indifference}
\end{align}
holds for $\theta=t_k(N)$. 
Indeed, if $\theta \in (t_{k-1}(N),t_{k}(N)]$, S may want to deviate to $m_{k+1}$, but it must be indifferent in equilibrium. 
Let $A=\frac{b^S_R+\sum_{w \in W}(b^S_w+b^w_R)}{|W|+1}$. From \eqref{eq:indifference} we have 
\begin{align*}
t_k(N)=k t_1(N) +2A k(k-1) \quad k=1, \dots, N.
\end{align*}
From the equations and $t_N(N)=1$, we see $t_1(N)=1/N-2A(N-1)$ and 
\[
\frac{1}{2N(N-k)} > A \quad \text{ for }k=1,\cdots, N-1.
\]
 It follows that 
\[A \leq \frac{1}{2N(N-1)}.\]
This forces 
\[
\frac{1}{2N(N+1)} \geq A
\], because if otherwise we see $N+1$ partitions in equilibrium. Hence, $b_{\mathrm{eff}}=A$. 
\end{proof}

\subsection*{Proof of Lemma \ref{lem:star-max}}
\begin{proof}
Fix a tree $T$ on $n$ nodes. 
Then we have
\[U_1(T)=n-1, \quad \sum_{t\geq 2}U_t(T)=\binom{n}{2}-(n-1)=\binom{n-1}{2}.\]
Hence
\[
v^G(S_{n-1}) =2 (n-1) \delta + 2 \binom{n-1}{2} \delta^2 \geq 2 (n-1) \delta + 2\sum_{t\geq 2}U_t(T)\delta^t  = v^G(T),
\] since $\delta^t \leq \delta^2$ for all $t \geq 2$ when $\delta \in (0,1)$. 
Equality holds iff $U_t(T)=0$ for all $t \geq 3$, that is, $\mathrm{diam}(T) \leq 2$, which characterizes star. 
\end{proof}

\subsection*{Proof of Proposition \ref{prop:single-witness}}
\begin{proof}
It is easy to see that the effective bias is   
 $b_{\mathrm{eff}}  =\delta+\delta^2$ regardless who is the hub.  
If there is no witness link, the bias is $b^S_R=\delta$. If the communication of S and R is privately done on the 3-node star, $b^S_R=\delta+2/3 \delta^2$.  In either case, the effective bias $b_{\mathrm{eff}}$ is larger.  
From (PT), the proposition statement holds. 
\end{proof}

\subsection*{Proof of Proposition \ref{prop:SorR-hub-case}}
\begin{proof}
Since  
\[
 b_{\mathrm{eff}}^{\mathrm{(\text{s.hub})}}
  = \delta\;+\;\underbrace{
      \frac{2(k+1)(k-1)}{3k}\,\delta^{2}
    }_{\textstyle f(k,\delta)}
\]
 strictly increasing in $k$  for all $\delta>0$, the number of partitions is non-increasing in $n=k+1$. 
Since $f(k,\delta)$, $\partial f/\partial k>0$ and $\partial^2 f/\partial k^2<0$, marginal change diminishes as $n \to \infty$. 
From (PT), the proposition statement holds. 
\end{proof}

\subsection*{Proof of Proposition \ref{prop:W-hub-case}}
\begin{proof}
From 
\[
  b_{\mathrm{eff}}^{\mathrm{(\text{w.hub})}}
  \;=\; 
\dfrac{2\delta+(4k-5)\tfrac23\delta^{2}}{\,k\,}=g(k, \delta), 
\] we have 
\[
\partial g/\partial k =\frac{-1}{k^2} \cdot 2 \delta \left(1-\frac{5}{3} \delta \right), \quad \partial^2 g/\partial k^2 =\frac{2}{k^3} \cdot 2 \delta \left(1-\frac{5}{3} \delta \right). 
\] It follows that $f\partial g/\partial k<0$, $\partial^2 g/\partial k^2>0$ for $\delta < 0.6$; $\partial g/\partial k \geq 0$, $ \partial^2 g/\partial k^2\leq 0$ for $\delta  \geq 0.6$. From (PT), the proposition statement holds. 
\end{proof}

\subsection*{Proof of Proposition \ref{prop:two-stars}}
\begin{proof}
Write
\[b_{\mathrm{eff}}^{\bigstar}:=b_{\mathrm{eff}}\big(S_m\text{ (hub}\to\text{leaf)}\big), \, b_{\mathrm{eff}}^{\star\star}:=
b_{\mathrm{eff}}\big(S_k\text{--}S_{\ell}\text{ (hub}\to\text{hub)}\big).\]

From \eqref{eq:eff_s_hub} we have 
\[
b_{\mathrm{eff}}^{\bigstar}=\delta + \frac{2}{3}\left( m-\frac{1}{m}\right)\delta^2.
\]
From \eqref{eq:eff} we know that 
\[
m \cdot b_{\mathrm{eff}^{\star\star}} = 2 v(S_k\text{--}S_{\ell})-v(S_k)-v(S_{\ell})-3\mu_S(S_k\text{--}S_{\ell};v)+2\mu_S(S_k;v)-\mu_R(S_{\ell};v). 
\]
From Lemma \ref{lem:tree-path-sharing-rule} we obtain 
\[
\mu_S(S_k\text{--}S_{\ell};v)=(k+1)\delta + \frac{k^2+k+2 \ell}{3}\delta^2+\frac{k\ell}{2}\delta^3. 
\]
It follows that  
\[
m \cdot b_{\mathrm{eff}^{\star\star}} = m\delta +\frac{2}{3}(m^2-1-2k\ell)\delta^2 + \frac{5}{2}k\ell \delta^3.
\]
Hence 
\[
b_{\mathrm{eff}}^{\bigstar}-b_{\mathrm{eff}}^{\star\star}=\frac{k\ell}{m}\delta^2 \left( \frac{4}{3}-\frac{5}{2}\delta \right).
\]
\end{proof}

\subsection*{Proof of Lemma \ref{lem:hub-hub-vs-leaf-leaf}}
\begin{proof}
We have 
\[
v(S_k\text{--}S_{\ell} \, \mathrm{hub}) = 2(k+\ell+1)\delta + 2\left(\binom{k}{2} +\binom{\ell}{2}+k+\ell \right)\delta^2 + 2k \ell \delta^3,
\] while
\begin{align*}
v(S_k\text{--}S_{\ell} \, \mathrm{leaf}) &= 
2(k+\ell+1) \delta + 2 \left(\binom{k}{2}+\binom{\ell}{2}+2 \right) \delta^2 \\ 
&+ 2(k+ \ell-1)\delta^3 + 2(k+\ell-2)\delta^4+2(k\ell -k-\ell+1)\delta^5.
\end{align*}
Since the sum of coefficients of $\delta^3$, $\delta^4$, and $\delta^5$ of the above equation is $2(k+\ell+k\ell-2)$, we see 
\[
v(S_k\text{--}S_{\ell} \, \mathrm{hub})>v(S_k\text{--}S_{\ell} \, \mathrm{leaf})
\] for $\delta \in (0,1)$. 
\end{proof}
\subsection*{Proof of Proposition \ref{prop:hub-hub-vs-leaf-leaf}}
\begin{proof}
Let S be a leaf of $S_k$ and R be a leaf of $S_\ell$, and S speaks to R.
Let $b_{\mathrm{eff}}^{\mathrm{hub}}$ and $b_{\mathrm{eff}}^{\mathrm{leaf}}$ be the effective bias for the case where the hubs communicate, and that for the case where the leaves communicate, respectively. 
From Proposition \ref{prop:two-stars} 
\[
b_{\mathrm{eff}}^{\mathrm{hub}}=\delta + \frac{2}{3}\left(m-\frac{1}{m}-\frac{2k\ell}{m} \right) \delta^2 +\frac{5 k \ell}{2m} \delta^3, 
\]
where $m=k+\ell+1$. 
Using the path-sharing rule, we obtain 
\[\mu_S(S_k\text{--}S_{\ell};v)=2\delta+\frac{2}{3}(k+1)\delta^2+\frac{1}{2}(k+\ell-1)\delta^3+\frac{2}{5}(k+\ell-2) + \frac{1}{3}(k-1)(\ell-1)\delta^3, \] 
\begin{align*}
b_{\mathrm{eff}}^{\mathrm{leaf}}&=
\frac{3}{m}\delta + \frac{4}{3}\delta^2 + \frac{5}{2}(m-2)\delta^3+\frac{14}{5}(m-3)\delta^4+3(k-1)(\ell-1)\delta^5,
\end{align*}
and 
\begin{align*}
\Delta(\delta) &= \left(1-\frac{3}{m}\right)\delta + \frac{2}{3} \left(m-\frac{1}{m}-\frac{2k \ell}{m}-2 \right) \delta^2 \\
&\hspace{12pt} \frac{5(k \ell -m +2)}{2m} \delta^3 -\frac{14(m-3)}{5m} \delta^4 -\frac{3(k-1)(\ell-1)}{m} \delta^5. 
\end{align*}
The first three terms on the left-hand side of the above equation are positive, and the last two terms are negative. 
Since $\delta \in (0,1)$, it is enough to check the sign of $\Delta(\delta)$ at $\delta=1$. 
The minimum point of $\Delta(1)$ is at $k=\ell=s^{\ast} \doteqdot 1.4823$ with $\Delta(1)^{\ast} \doteqdot -0.06465$. 
Since $\Delta(\delta)$ is convex, symmetric, and U-shaped in each coordinate at the balanced point  $k=\ell=s^{\ast} \doteqdot 1.4823$, this yields:
\begin{itemize}
\item $\Delta(1)=0$ at $(1,1)$
\item $\Delta(1)=-0.02$ at $(2,2)$
\item $\Delta(1)>0$ everywhere else.
\end{itemize}
For $k=\ell=2$, $\Delta(\delta)$ crosses zero at 
\[
\delta_c \doteqdot 0.9949.
\]
\end{proof}
\end{appendices}
\end{document}